\documentclass[acmtog]{acmart}
\usepackage{booktabs} 

\usepackage[ruled]{algorithm2e} 
\usepackage{xcolor}
\usepackage{placeins}
\usepackage{graphicx}
\usepackage{hyperref}

\SetAlFnt{\small}
\SetAlCapFnt{\small}
\SetAlCapNameFnt{\small}
\SetAlCapHSkip{0pt}

\acmJournal{TOG}

\copyrightyear{2024} \acmYear{2024} \setcopyright{acmlicensed}\acmConference[SA Conference Papers '24]{SIGGRAPH Asia 2024 Conference Papers}{December 3--6, 2024}{Tokyo, Japan}\acmBooktitle{SIGGRAPH Asia 2024 Conference Papers (SA Conference Papers '24), December 3--6, 2024, Tokyo, Japan}\acmDOI{10.1145/3680528.3687669}\acmISBN{979-8-4007-1131-2/24/12}

\begin{document}
\title{FreeAvatar: Robust 3D Facial Animation Transfer by Learning an Expression Foundation Model}

\begin{abstract}

Video-driven 3D facial animation transfer aims to drive avatars to reproduce the expressions of actors. Existing methods have achieved remarkable results by constraining both geometric and perceptual consistency. However, geometric constraints (like those designed on facial landmarks) are insufficient to capture subtle emotions, while expression features trained on classification tasks lack fine granularity for complex emotions. To address this, we propose \textbf{FreeAvatar}, a robust facial animation transfer method that relies solely on our learned expression representation. Specifically, FreeAvatar consists of two main components: the expression foundation model and the facial animation transfer model. In the first component, we initially construct a facial feature space through a face reconstruction task and then optimize the expression feature space by exploring the similarities among different expressions. Benefiting from training on the amounts of unlabeled facial images and re-collected expression comparison dataset, our model adapts freely and effectively to any in-the-wild input facial images.
In the facial animation transfer component, we propose a novel Expression-driven Multi-avatar Animator, which first maps expressive semantics to the facial control parameters of 3D avatars and then imposes perceptual constraints between the input and output images to maintain expression consistency. To make the entire process differentiable, we employ a trained neural renderer to translate rig parameters into corresponding images. Furthermore, unlike previous methods that require separate decoders for each avatar, we propose a dynamic identity injection module that allows for the joint training of multiple avatars within a single network. The comparisons show that our method achieves prominent performance even without introducing any geometric constraints, highlighting the robustness of our FreeAvatar. Our code will be publicly available at \textcolor{blue}{\href{https://github.com/FuxiVirtualHuman/free_avatar}{here}}.

\end{abstract}



\author{Feng Qiu}
\affiliation{%
 \institution{NetEase Fuxi AI Lab}
 \city{Hangzhou}
 \state{Zhejiang}
 \country{China}}
\email{qiufeng@corp.netease.com}

\author{Wei Zhang}
\affiliation{%
 \institution{NetEase Fuxi AI Lab}
 \city{Hangzhou}
 \state{Zhejiang}
 \country{China}}
\email{zhangwei05@corp.netease.com}

\author{Chen Liu}
\affiliation{%
 \institution{The University of Queensland}
 \city{Brisbane}
 \state{Queensland}
 \country{Australia}
}
\email{chen.liu7@uqconnect.edu.au}

\author{Rudong An}
\affiliation{%
 \institution{NetEase Fuxi AI Lab}
 \city{Hangzhou}
 \state{Zhejiang}
 \country{China}}
\email{anrudong@corp.netease.com}

\author{Lincheng Li}\thanks{*Corresponding author: Lincheng Li, lilincheng@corp.netease.com}
\affiliation{%
 \institution{NetEase Fuxi AI Lab}
 \city{Hangzhou}
 \state{Zhejiang}
 \country{China}}
\email{lilincheng@corp.netease.com}

\author{Yu Ding}
\affiliation{%
 \institution{NetEase Fuxi AI Lab}
 \city{Hangzhou}
 \state{Zhejiang}
 \country{China}}
\email{dingyu01@corp.netease.com}

\author{Changjie Fan}
\affiliation{%
 \institution{NetEase Fuxi AI Lab}
 \city{Hangzhou}
 \state{Zhejiang}
 \country{China}}
\email{fanchangjie@corp.netease.com}

\author{Zhipeng Hu}
\affiliation{%
 \institution{NetEase Fuxi AI Lab}
 \city{Hangzhou}
 \state{Zhejiang}
 \country{China}}
\email{zphu@corp.netease.com}

\author{Xin Yu}
\affiliation{%
 \institution{The University of Queensland}
 \city{Brisbane}
 \state{Queensland}
 \country{Australia}
}
\email{xin.yu@uq.edu.au}


\renewcommand\shortauthors{F. Qiu, W. Zhang, C. Liu, R. An, L. Li, Y. Ding, C. Fan, Z. Hu, X. Yu}

%
%




\begin{teaserfigure}
\centering
  \includegraphics[width=1\textwidth]{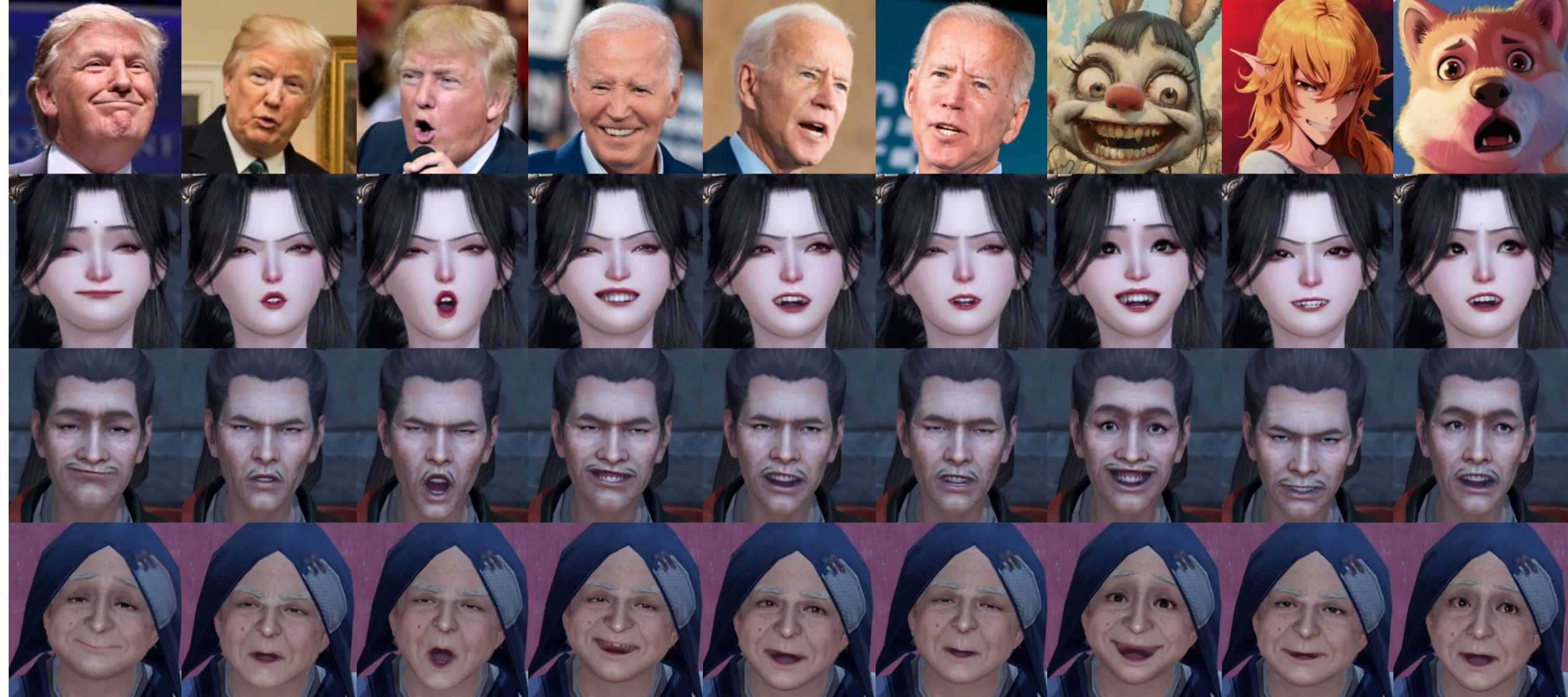}
  \caption{\textbf{FreeAvatar} is capable of driving different 3D avatars to generate high-fidelity facial animations consistent with the expressions in the input in-the-wild images. The first row shows the input facial images with compound expressions. The following rows are the results based on different avatars. 
  First and third images of \textit{Donald Trump} \textcopyright{Gage Skidmore/Flickr} (CC BY-SA 2.0). Second image of \textit{Donald Trump} \textcopyright{Trump White House Archived/Flickr} (public domain). Images of \textit{Joe Biden} \textcopyright{Joe Biden/Flickr} (CC BY-NC-SA 2.0). \textit{Cartoon characters} \textcopyright{NetEase}.}
  \label{fig:teaser}
\end{teaserfigure}
\maketitle

\section{Introduction}
\label{sec:intro}


3D facial animation transfer methods aim to capture human facial expressions and movements to create realistic animations for digital avatars, which have vast application prospects in digital human, CG games, VR and AR, etc. \cite{nowak2018avatars, davis2009avatars, zollhofer2018state}. 
Facial motion capture systems, such as Faceware~\cite{faceware} and ARKit~\cite{arkit}, are extensively utilized in practical applications~\cite{cao2013facewarehouse, furukawa2009dense}. 
Compared to manually created animations, it can present more delicate facial expressions. With the development of computer vision, video-driven facial animation transfer methods have gained considerable attention due to their convenience and low cost. However, achieving natural and accurate expression transfer while ensuring consistency in facial emotions is still a challenge.

Existing methods \cite{aneja2018learning, zhang2020facial, pan2023real} simultaneously employ facial geometry priors and expression features to maintain emotional semantic consistency between source and target faces. 
We find that these methods often fail to drive the target avatar to generate high-fidelity expressions, especially for in-the-wild data.
There are several reasons for this. 
Firstly, the geometric constraints, like metrics based on facial landmarks, struggle to effectively capture subtle changes in expressions, such as slight frowning and the compression of lips. Secondly, the expression features are commonly trained on the discrete emotion classification task based on limited categories. However, human emotional variations are often diverse and continuous. Therefore, these expression features are unable to capture fine-grained emotional differences.

To tackle the aforementioned issues, we introduce \textbf{FreeAvatar}, a robust facial animation method that relies solely on the expression representation and is capable of maintaining high-fidelity expressions.
The core idea of FreeAvatar is first to learn a continuous and semantic distinguishable expression representation that can be derived from any facial image and then devise an animation transfer model that can precisely decode the expression representation into the target avatar expression.

We begin by learning an expression foundation model to construct a fine-grained and expressive latent space. Within this space, facial images with similar expressions cluster together, while those with dissimilar expressions distance themselves apart. 
Specifically, we first utilize the Masked Autoencoder (MAE)~\cite{he2022masked} to learn the intrinsic facial features from a large amount of unlabeled facial images. 
Based on the pre-trained ViT encoder, we then incorporate contrastive learning to finetune it. 
Unlike previous works~\cite{zhang2020facial, larey2023facial, pan2023real} that constrain the expression consistency by coarse features trained on a limited set of discrete categories, our method mimics human perception by learning to recognize the subtle nuances of facial expressions. This approach allows the model to achieve a fine-grained and continuous expression feature.
After that, we attain the powerful expression foundation model.

Afterward, we propose a novel Expression-driven Multi-avatar Animator to produce facial animations from the extracted expression representations. In this component, we first leverage a rig parameter encoder to map the expression representation into the facial control parameters of 3D avatars. 

To capture high-frequency details of emotions, we then employ a neural renderer to translate these parameters into the facial images of target avatars. 
Considering that previous works require training separate decoders for each target avatar, which significantly limits model scalability, we also propose a dynamic identity injection module. This module enables joint training of multiple avatars by randomly assigning the avatar ID during training. Additionally, to enhance the model's generalizability, we devise an identity-conditional loss to achieve semi-supervised training. This loss functions by enforcing constraints between facial control parameters and image pixels only when the IDs of the input and output match.

Extensive experiments demonstrate our FreeAvatar can achieve state-of-the-art performance from in-the-wild images, which allows our method to be freely and conveniently applied to various scenarios.

In summary, the contributions of this work are listed as:

\begin{itemize}
    \item We propose FreeAvatar, the first method that relies solely on expression representations for 3D facial animation transfer. Leveraging our expression triplet dataset, this approach enables high-fidelity 3D facial animation transfer, even with in-the-wild face images.
    \vspace{0.3em}
    \item We introduce an expression foundation model designed to construct a universal, fine-grained, and continuous latent space that adapts well to various faces, including stylized avatars. Benefiting from this model, FreeAvatar can maintain a high level of expression consistency during facial animation transfer.
    \vspace{0.3em}
     \item  We devise an Expression-driven Multi-avatar Animator to decode expression representations into facial control parameters and maintain expression consistency. The dynamic identity injection module and the identity-conditional loss allow us to implement animations for multiple avatars with only one decoder.
    \vspace{0.3em}

\end{itemize}

\section{Related Work}
\label{sec:relatedworks}
\subsection{3D Facial Animation Transfer}
In the film and gaming industries, facial animations are mainly controlled by facial rigs using a method called blendshape-based animation \cite{lewis2014practice}. The process of 3D facial animation transfer involves replicating the performer's facial movements accurately by capturing the blendshape weights to manipulate the 3D face, which may not have the same physiognomy as the performer.
While facial motion capture systems such as Faceware~\cite{faceware}, ARKit~\cite{arkit}, and Metahuman Animator~\cite{metahuman_animator} enhance the naturalness and realism of animations, these approaches are typically expensive, hardware-dependent, and necessitate intricate calibration processes.


Recent advances in computer vision have led to two main methods for 3D facial animation transfer.
The first method involves obtaining actor-specific 3D facial animations through monocular face reconstruction technology~\cite{danvevcek2022emoca, tewari2017mofa, feng2021learning,wang2022faceverse,lei2023hierarchical}, which are then retargeted for the target avatars~\cite{ribera2017facial, pighin2006facial, chandran2022local}.  
While these approaches perform well, they are limited by the requirements of manually set up mappings for expression semantic or facial geometric, which reduces their flexibility and scalability.
The second method involves using neural networks to directly produce the 3D control parameters of target avatars from source facial images.
For instance, ExprGen \cite{aneja2018learning} learns to correlate 2D images with avatar expressions via perceptual and facial geometric constraints.
Animatomy~\cite{choi2022animatomy} introduces muscle fiber curves to build a modular deformation system.
\citet{larey2023facial} propose a deep-learning architecture that en
\citet{larey2023facial} propose a deep-learning architecture that encodes the landmarks of each facial organ to the blendshape weights of target avatars.
\citet{pan2023real} presents a blendshape adaption network that maps the source facial images to the rig parameters of target avatars by minimizing the geometric and emotional distances. Unlike previous work using both geometric and perception constraints, the work by \citet{moser2021semi} is the only one that exclusively utilizes the expression representation for 3D facial animation transfer.
This approach involves training separate image-to-image models first and then image-to-geometry models. 

Despite the above methods simplifying the process of animation transfer, these methods struggle to maintain emotional semantic consistency, especially for in-the-wild data.
In contrast,  our FreeAvatar eliminates geometric priors and constructs a fine-grained, continuous expression representation to produce facial animations with high-fidelity expressions for multiple avatars. 

\subsection{Facial Expression Representation}
Emotion analysis within the realm of deep learning has seen considerable achievements over the years, driven by the requirements of understanding and replicating human emotions in machines \cite{scherer2010blueprint, poria2017review, hakak2017emotion,danvevcek2023emotional}.
Typical methods utilize discrete emotion categories to represent facial expressions, including basic emotions outlined by \citet{ekman1992argument} and \citet{kollias2022abaw} and compound emotions discussed by \citet{ekmanp1978tech} and \citet{shao2018deep}.
The Facial Action Coding System and Action Unit (AU) proposed by \citet{ekmanp1978tech} and enhanced by \citet{shao2018deep} focuses on identifying specific facial muscle movements related to expressions.
Despite their widespread adoption, these discrete categories usually fail to capture the nuanced aspects of facial expressions. 
Several studies~\cite{drobyshev2022megaportraits,trevithick2023real} predict canonical triplane representations from images to extract facial features. 
However, these methods struggle to disentangle expressions from appearance effectively. In addition, speech-driven 3D avatar animation approaches~\cite{aneja2024facetalk,fan2022faceformer,karras2017audio} are limited to dialogue or speech scenarios, which renders them unsuitable for non-verbal contexts.

To address this problem, continuous Valence and Arousal (VA) \cite{russell1989cross} along with the emotion intensity \cite{kollias2022abaw} provide a new way to give the fine-grained emotion representation.
3D morphable models~\cite{chandran2020semantic,tran2019towards,paysan20093d,FLAME:SiggraphAsia2017,tewari2021learning} offer a robust framework for representing and analyzing 3D facial characteristics.
However, these approaches rely on high-cost training data. Therefore, \citet{zhang2021learning} introduces an identity-invariant expression embedding space for expression recognition. 
However, since the work focuses on eliminating the impacts of identities, the space is built on the real-human dataset and cannot handle the stylized cartoon characters. Additionally, the expression space learned in this work overlooks the asymmetry of expressions. 
In contrast, FreeAvatar leverages a re-collected and more extensive dataset to reconstruct a more expressive and comprehensive expression space for 3D facial animation transfer.

\begin{figure*}[t]
  \centering
  \includegraphics[width=1\linewidth]{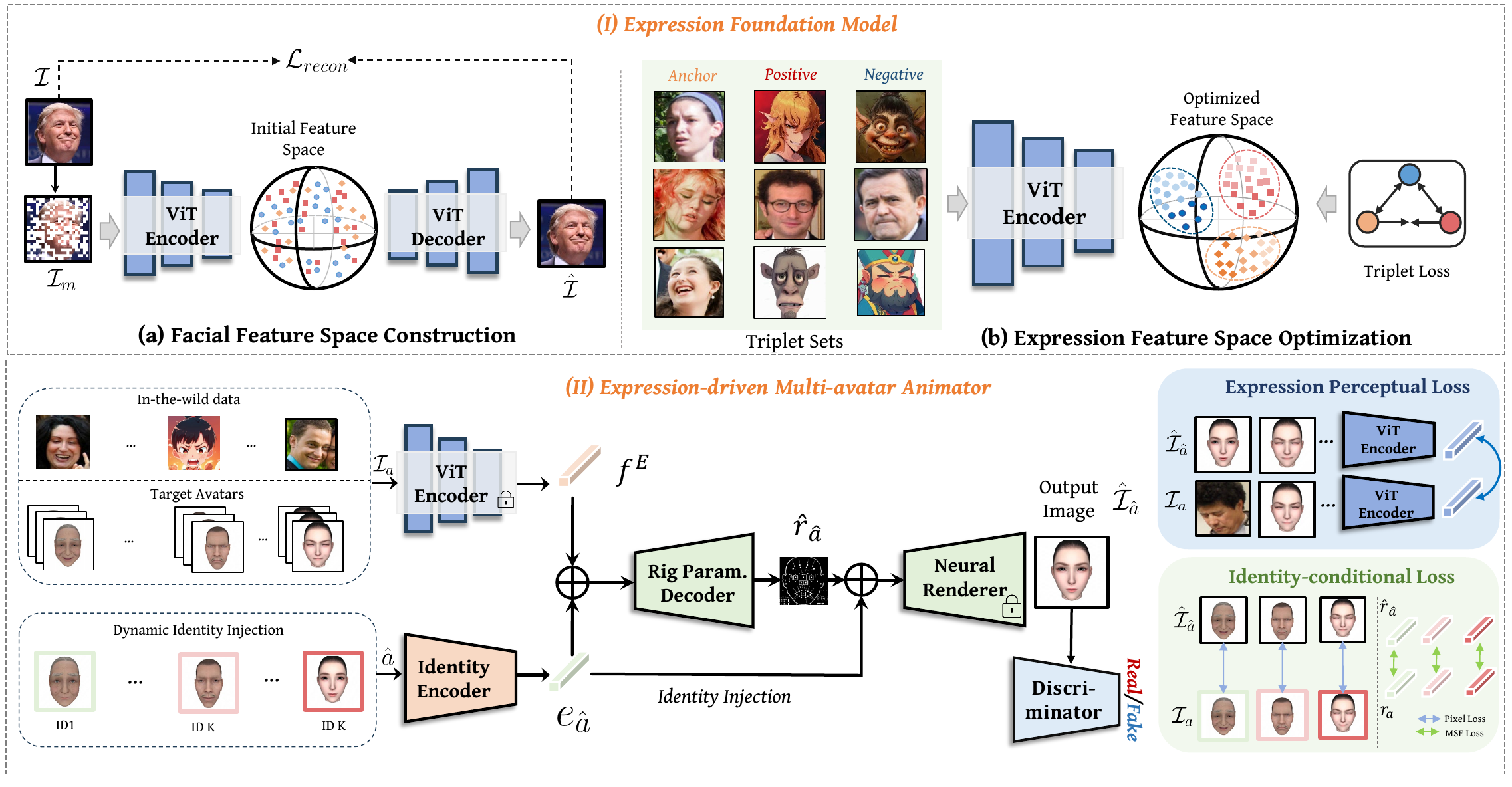}
  \caption{\textbf{Pipeline overview}. FreeAvatar first constructs an expression foundation model in two steps: facial feature space construction with Masked Autoencoder (MAE) and expression feature space optimization via contrastive learning. After that, an Expression-driven Multi-avatar Animator is constructed 
  to encode the expression representations into rig parameters. Then, perceptual constraints are employed in a differentiable manner to ensure that the expressions between the input and the avatars remain consistent.}
  \label{pic:pipeline}
\end{figure*}

\section{METHOD}
Existing 3D facial animation transfer methods \cite{moser2021semi, larey2023facial, pan2023real} leverage facial geometric information and expression representation to ensure consistency of expression between input and output images. However, as mentioned before, existing constraints are not effective at capturing the subtle details of facial expression contents.
In our FreeAvatar, we introduce a novel expression representation to capture subtle expressions for high-fidelity facial animation transfer. To achieve this, as depicted in Fig.~\ref{pic:pipeline}, we first construct an expression foundation model in two steps: facial feature space construction and expression feature space optimization. Afterward, we propose an Expression-driven Multi-avatar Animator that achieves high-fidelity facial animation transfer and adapts effectively to in-the-wild facial images.



\subsection{Expression Foundation Model}
\label{sec:med_ExprFoundation}

Previous works \cite{pan2023real, danvevcek2022emoca} typically derive the expression extractor via implementing an emotion recognition task.
However, we observe that this approach has several limitations:
(i) the limited set of expression categories cannot cover all possible expression variations, 
(ii) the discrete predictions fail to capture the subtle differences between emotions within the same category, 
(iii) existing datasets only contain real human images, leading to a domain gap between them and stylized avatars, which can result in poor generalization of the extracted features.

To tackle these issues, we incorporate contrastive learning to build an expression foundation model, which captures fine-grained and continuous emotion features. To enhance the model's generalization capabilities, we initially use the masked autoencoder (MAE)~\cite{he2022masked} to develop a robust facial feature extractor through a self-reconstruction task. Leveraging the pretrained ViT encoder with MAE, we then fine-tune this model on the re-collected expression comparison dataset to attain our expression foundation model.

\subsubsection{Facial feature space construction.} To effectively utilize a large amount of unlabeled facial images, we first employ MAE pre-training to learn the intrinsic features and structure of faces, thereby enhancing the model's generalization capability. 
Given a facial image $\mathcal{I}$, we divide it into 16 $\times$ 16 non-overlapping patches. Then we mask 75\% of the image area as \citet{he2022masked} and obtain $\mathcal{I}_{m}$.
After that, we employ a ViT encoder $\textit{E}$ to encode the $\mathcal{I}_{m}$ into a latent feature $f^E$ and a ViT decoder $\textit{D}$ to decode the representation into the original portrait image.
For learning of the facial feature space, we optimize the parameters of the autoencoder by the L2 loss.
Afterward, the ViT encoder $\textit{E}$ can serve as a powerful facial feature extractor.


\subsubsection{Expression feature space optimization.} 
Based on the pretrained encoder $\textit{E}$, we fine-tune the model on an expression comparison dataset and optimize the latent space of expression. 
Specifically, given an expression triplet $\{\mathcal{I}_a, \mathcal{I}_p, \mathcal{I}_n\}$ with comparison annotation, we first use the encoder $\textit{E}$ to map them into the same latent space. 
\begin{equation}
    f_a = E(\mathcal{I}_a), f_p = E(\mathcal{I}_p), \text{and }  f_n = E(\mathcal{I}_n),
\end{equation}
where $a, p, n$ refer to \textit{anchor}, \textit{positive} and \textit{negative} in a triplet, and compared to $\mathcal{I}_n$, the facial expressions of $\mathcal{I}_a$ and $\mathcal{I}_p$ are more similar. During training, we use triplet loss to ensure that the representation distance between $\mathcal{I}_a$ and $\mathcal{I}_p$ is greater than that between $\mathcal{I}_a$ and $\mathcal{I}_n$. In other words, within the expression latent space, we pull $f_a$ and $f_p$ closer together and push away $f_a$ from $f_n$.

The expression feature space is optimized by the weighted triplet loss $\mathcal{L}_{tri}$, which can be formulated as:
\begin{equation}
    \mathcal{L}_{t r i}= w \cdot \textit{Max} \left(0,\left\|f_a-f_p\right\|_2-\left\|f_a-f_n\right\|_2+m\right).
\end{equation}
Here, $w$ indicates the confidence score, calculated by the ratio of agreed annotations to the total annotations of a sample. $m$ is the margin to ensure the anchor and positive images are closer together in the latent space than the anchor and negative images.
Once trained, the encoder $\textit{E}$ can serve as the expression foundation model. 


\subsection{Expression-driven Multi-avatar Animator}
\label{sec:expr_transfer}
Based on the feature extracted by our expression foundation model, we are able to drive avatars to produce corresponding facial animations. 
Unlike previous works that learned in an avatar-specific manner, our method achieves multi-avatar facial transfer within a single network. Specifically, we first devise a dynamic identity injection module that allows for the joint training of multiple avatars. Following this, we train a rig parameter decoder $\mathcal{R}$ to map the expression representation into the facial rigs of 3D avatars and employ perceptual constraints to ensure the consistency of the transferred expressions. 
In order to make the entire training process differentiable, we use a neural renderer to translate the rig parameters $\mathcal{N}$ into facial images of the target avatar.

\subsubsection{Expression feature extraction}
Based on our expression foundation model $E$, we are able to obtain the expression representation $\mathbf{f}$ from the source facial image $\mathcal{I}_a$.
\begin{equation}
        \mathbf{f} = E(\mathcal{I}_a), \textit{a} \in \{0, 1, \dots, K\},
\end{equation}
where $\textit{a}$ indicates the identity of the input face. To enhance the model's generalizability, the training data includes not only target avatar images paired with rigs $\mathbf{r}_a$ but also unlabeled in-the-wild facial data (e.g., real human images or other stylized cartoon character images).
Specifically, $a \in \{ 1, \dots, K\}$ refers to the identity number of the target avatars, and $a=0$ indicates that the data is an in-the-wild facial image. And $K$ is the total number of the target avatars.

\subsubsection{Dynamic identity injection.}
To accomplish joint training for multiple avatars, during the training process, we randomly assign the target avatar and dynamically inject them into the rig decoder $\mathcal{R}$ and the neural renderer $\mathcal{N}$. Specifically, for each iteration, we randomly choose $\hat{a} \in \{1,..., K\}$, which indicates the identity number of the target avatar. Then we employ an Embedding Layer as the identity encoder $\textit{E}_{ID}$ to extract identity embedding $e_{\hat{a}}=\textit{E}_{ID}(\hat{a})$.



\subsubsection{Rig parameter decoder}
After that, we should map the expressive semantic information to the facial controllers of the 3D avatar.
To achieve this, we devise the Rig Parameter Decoder $\mathcal{R}$ consisting of Multi-Perceptron Layers (MLPs).
Specifically, it decodes the expression representation $\mathbf{f}$ into rig parameters of the target avatar.
Since different avatars possess varying rigs and unique physiognomy, the generated rig parameters of different avatars not only need to contain consistent expression information but also possess unique facial attributes. 
Hence, in the decoding process, we also incorporate the identity embedding $\hat{a}$ and concatenate it to the expression representation.
The whole process is expressed by:
\begin{equation}
    \mathbf{\hat{r}}_{\hat{a}} = \mathcal{R}(\mathbf{f}, e_{\hat{a}}), \hat{a} \in \{1, \dots, K\}.
\end{equation}

\subsubsection{Neural renderer}
After that, the rig parameters should be translated into the facial images of target avatars to enable expression supervision and capture high-frequency details during training. To make this process differentiable, we employ a pre-trained neural renderer $\mathcal{N}$ to mimic the 3D render engine. Specifically, we take the architecture of DCGAN generator~\cite{radford2015unsupervised} as the backbone of $\mathcal{N}$ and train it on the rig-image paired dataset. It takes the rigs as input and produces the corresponding avatar images.
Moreover, to ensure the identity consistency of avatars, we also add the encoded identity information to the neural renderer, formulated as:
\begin{equation}
    \hat{\mathcal{I}}_{\hat{a}} = \mathcal{N}(\mathbf{\hat{r}}_{\hat{a}}, e_{\hat{a}}),
\end{equation}
where $\hat{\mathcal{I}_{\hat{a}}}$ refers to the rendered image of avatar 
 $\hat{a}$.

\subsubsection{Training objectives}

As mentioned before, the training data consists of two parts, and the in-the-wild facial images have no ground truth for rig parameters or rendered images of target avatars. Therefore, our pipeline is trained in a semi-supervised manner.

\paragraph{Perception loss:}
First, we extract expression embeddings from the output image $\hat{\mathcal{I}}_{\hat{a}}$ and the input image ${\mathcal{I}}_{{a}}$ and encourage consistency between them and obtain expression perception loss:
\begin{equation}
    \mathcal{L}_{expr} = ||\mathbf{f} - \textit{E}(\hat{\mathcal{I}}_{\hat{a}}) ||^2.
\end{equation}

\paragraph{Generative adversial loss:}
To enhance the quality of the generated rig parameters and make them as close as possible to real data, our framework also incorporates a Discriminator $\mathcal{D}$ to form a Generative Adversarial Network (GAN). The generative adversarial loss can be formulated as follows:
\begin{equation}
\mathcal{L}_{\mathrm{GAN}}=\mathbb{E}_{\mathcal{I}_{a}}\left[\log \mathcal{D}\left(\mathcal{I}_{a}\right)\right]+\mathbb{E}_{\hat{\mathcal{I}}_{\hat{a}}}\left[\log \left(1-\mathcal{D}\left(\hat{\mathcal{I}}_{\hat{a}}\right)\right)\right].
\end{equation}

\paragraph{Cycle loss:}
To enhance the model's generalization ability, we further introduce a cycle consistency loss that transfers the generated expressions of $\hat{\mathcal{I}}_{\hat{a}}$ back onto the target character. This practice can reduce the domain gap between target avatars and in-the-wild faces, thereby improving the generalization to unseen facial images. The cycle consistency loss can be formulated as follows:
\begin{equation}
    \mathcal{L}_{cycle} =  || \mathbf{\hat{r}}_{\hat{a}} - \mathcal{R}(\textit{E}(\hat{\mathcal{I}}_{\hat{a}}), \hat{a})||^2.
\end{equation}

\paragraph{Identity-conditional loss:}
In this work, we curate two types of facial image datasets: one comprising avatar images paired with rig parameters $\mathbf{r}_a$, and another without.
Considering leveraging paired data in training can enhance the model's generalization and performance, we propose an identity-conditional loss within a semi-supervised learning framework.
During training, this constraint is applied only to the image data with paired rig parameters. Specifically, when the target avatar identity $\hat{a}$ matches the identity $a$ from the input image, the rig-image pairs can be used to enhance the network's accuracy and accelerate convergence. Otherwise, this supervision is not performed. Therefore, the identity-conditional loss $\mathcal{L}_{IDC}$ can be formulated as:


\begin{equation}
    \mathcal{L}_{IDC} = \begin{cases} 
    ||\mathcal{I}_{a} - \hat{\mathcal{I}}_{\hat{a}}||^2 + ||\mathbf{r}_{a} - \hat{\mathbf{r}}_{\hat{a}}||^2, & \text{if } a = \hat{a}, \\
    0, & \text{otherwise}.
    \end{cases}
\end{equation}


\paragraph{Total loss:}
In summary, the final loss can be defined as
\begin{equation}
    \mathcal{L} = \lambda_{1}\mathcal{L}_{expr} + \lambda_{2}\mathcal{L}_{GAN} + \lambda_{3}\mathcal{L}_{\mathrm{cycle}} + \lambda_{4}\mathcal{L}_{IDC},
\end{equation}
where $\lambda_1,\lambda_2,\lambda_3,\lambda_4$ are the weights of different loss. And we set $\lambda_1=100$, $\lambda_2=\lambda_3=1e-3$ and $\lambda_4=1$ in this work.

\section{Experiments}
\label{sec:exp}

In this section, we assess the effectiveness of our framework by extensive experiments. Given a fully rigged Avatar model, our method achieves 3D facial animation transfer with only RGB images for in-the-wild scenarios.

\subsection{Data Acquisition}
\subsubsection{Unlabled facial images.} To enhance the generalization ability of our feature extractor, we integrate five facial databases (\emph{i.e.,} AffectNet \cite{mollahosseini2017affectnet}, CASIA-WebFace \cite{CASIA-Webface}, CelebA \cite{CelebA}, IMDB-WIKI \cite{IMDB-WIKI}, and WebFace260M \cite{zhu2021webface260m}) to form a large-scale facial dataset. Additionally, considering the gap between avatars and real humans, we also gather a part of stylized facial images from cartoons and animations as a supplement. Our dataset includes various races and genders. 
In total, this dataset comprises around 4.5 million facial images. 

\begin{figure*}[t]
  \centering
  \includegraphics[width=1\linewidth]{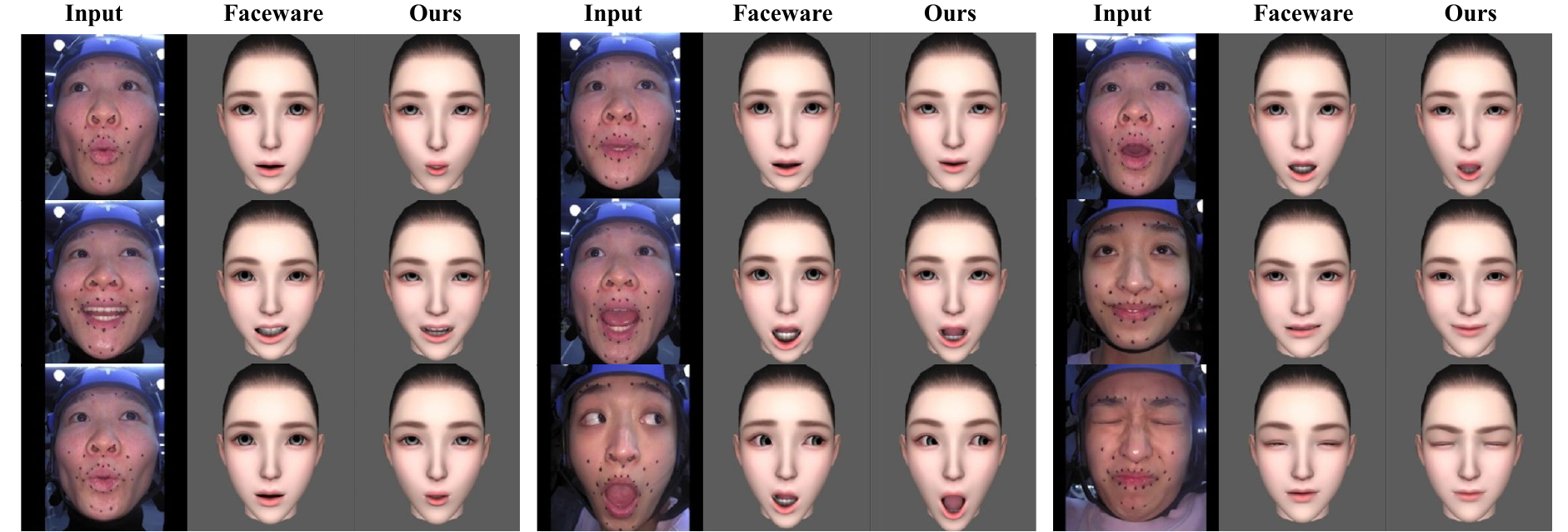}
  \caption{Comparisons with Faceware. Our method captures more detailed expressions, particularly the mouth movements during dialogue.}
  \label{pic:exp_faceware}
\end{figure*}

\subsubsection{Labeled expression triplets.}
\label{sec:triplet_data}
To enable the expression latent space to be continuous and expressive, we construct a multi-identity dataset consisting of 914K face image triplets with the expression comparison annotations. These triplets consist of two parts: one part ($\sim$500K triplets) is data with a high ratio of agreement from the FEC database ~\cite{vemulapalli2019compact}, and the other ($\sim$414K triplets) is randomly constructed from unlabeled facial images, involving real humans and cartoon images, which are then annotated. The scarcity of asymmetrical expression samples in the FEC dataset makes a trained model fail to represent facial asymmetry effectively. Therefore, we supplement the dataset with additional images of asymmetrical expressions. To maintain the labeling consistency, annotators are guided to focus on overall facial expressions rather than specific facial features. Moreover, we discard triplets if over 40\% of annotators cannot reach a consensus.

\subsubsection{Rig-image pairs.}
\label{sec:animi_data}
To facilitate the facial animation transfer, we build a dataset consisting of facial rig parameters paired with corresponding avatar facial images. 
Notably, our method does not necessitate temporally continuous samples; thus, we obtain a set of rig parameters via random sampling. Then these rigs are rendered into the facial images with a fixed front-facing head pose and a fixed camera setup.
In our experiments, we collect data from four different avatars, amassing 100K pairs of images and rig parameters for each avatar.


\subsection{Experimental Setup}
We implement the proposed approach in PyTorch and adopt AdamW optimizer~\cite{loshchilov2017decoupled}. For the facial feature space construction, training occurs over 800 epochs with a learning rate of 1e-4. We use a batch size of 4096, and the training is performed on 8 NVIDIA A30 GPUs. For the expression feature space optimization, the training spans 150 epochs with a learning rate of 0.002 and a batch size of 128. 
The training for facial animation transfer is conducted for up to 100,000 steps, employing a batch size of 16 and a learning rate set to 1e-5. Besides, we incorporate 200K unlabeled facial images along with the rig-image pairs to train our model.


\subsection{Comparisons with Commercial Products}
To assess the effectiveness of our pipeline, we first compare our FreeAvatar with the commercial facial motion capture systems, i.e. Faceware and MetaHuman Animator, which are the only viable and equivalent systems to our pipeline. These methods only adapt well to data collected under controlled conditions.

\subsubsection{Comparison with Faceware.}
In this scenario, we compare our method with Faceware on the frontal facial images from two actors.
Qualitative results can be seen in Fig.\ref{pic:exp_faceware}. It shows that our method excels in capturing the details of mouth movements during dialogue compared with Faceware. Specifically, Faceware frequently fails to capture key mouth movements along the viewing axis, such as puckering, which are vital for adding realism to the mouth animations. In contrast, thanks to the powerful expressive capabilities of our FreeAvatar, which utilizes only frontal view images, it effectively transfers such nuanced lip motions. This clearly demonstrates that the animations generated by our FreeAvatar not only replicate the facial expressions from the input images more accurately but also perceive the subtle differences between various facial actions, affirming its superior performance.


\begin{figure*}[t]
  \centering
  \includegraphics[width=0.98\linewidth]{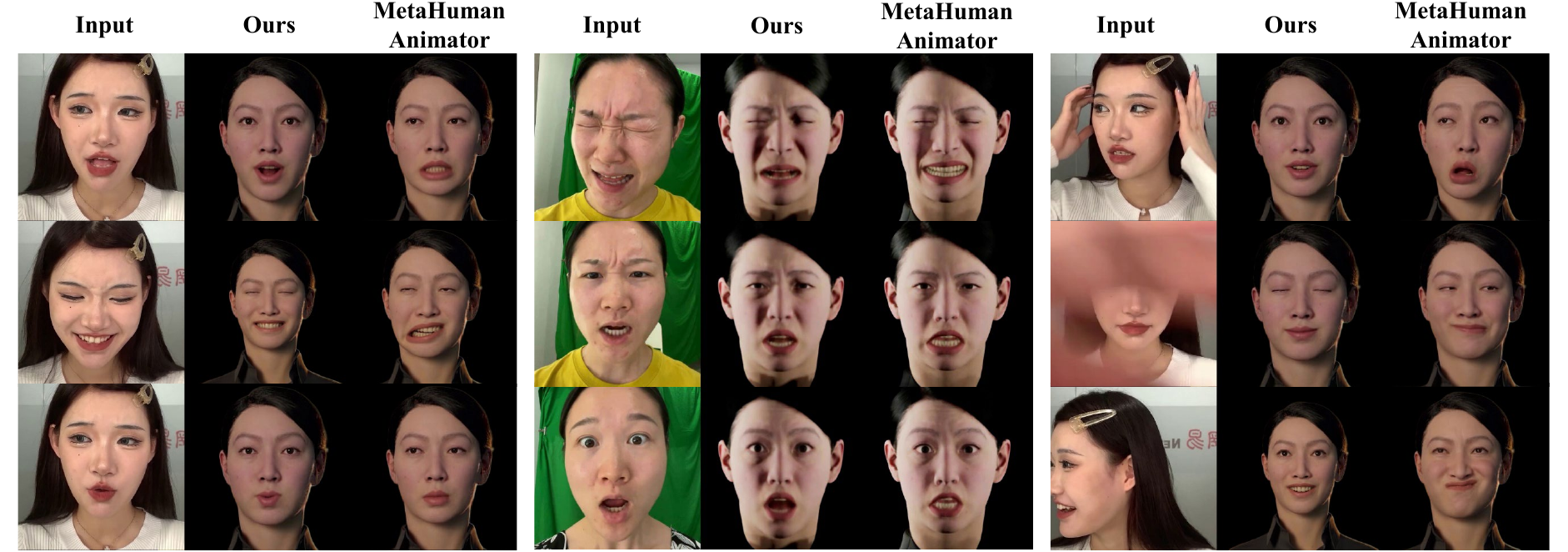}
  \caption{Comparisons with MetaHuman Animator. Our FreeAvatar demonstrates enhanced robustness in non-ideal conditions.}
  \label{pic:exp_animator}
\end{figure*}

\begin{figure}[!h]
  \centering
  \includegraphics[width=0.98\linewidth]{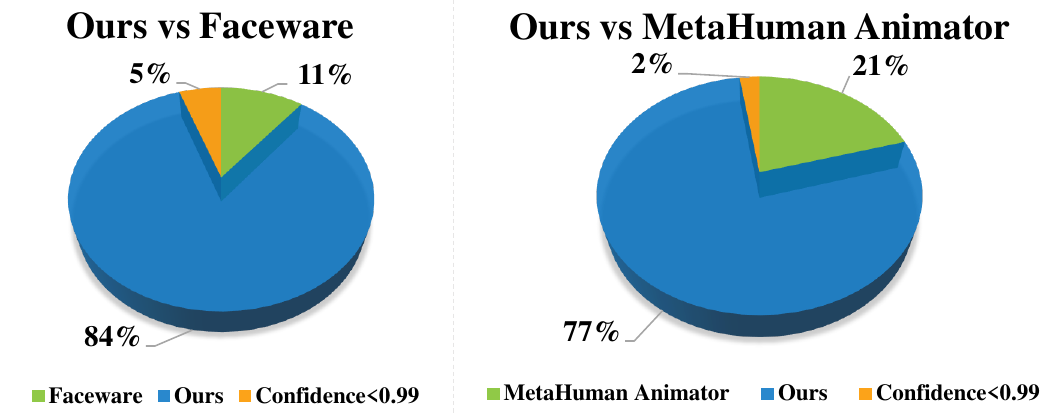}
  \caption{Distribution of user study results comparing the expression consistency between our method and commercial methods, Faceware and MetaHuman Animator. The results demonstrate that our FreeAvatar significantly outperforms the others.}
  \label{pic:user_study}
\end{figure}

\subsubsection{Comparison with MetaHuman Animator.}
MetaHuman Animator is a framework proposed by EPIC Games to create and animate highly realistic digital human characters. 
We use an iPhone to acquire the data and import them into the MetaHuman Animator for solving to obtain facial animation. 
As indicated in Fig.~\ref{pic:exp_animator}, our results are significantly better in the lip region than those of the MetaHuman Animator, even though it requires depth information. Additionally, the MetaHuman Animator demands stringent conditions for data collection. Minor changes in head posture, background variations, and obstructions can severely degrade the quality of the animations. Differently, our framework exhibits good robustness. Even when parts of the face are obscured, our approach is able to stably transfer facial movements of the visible parts, demonstrating strong robustness. 


\subsubsection{User study.}
In addition, we conduct a user study to evaluate the performance of expression transfer in our framework. A total of 21,451 survey questions were administered to over 39 participants. These questions are divided into two categories: 8,332 questions comparing our method with Faceware, and the remainder comparing our method with MetaHuman Animator. Each question presents three images: one input facial image and two 3D avatar images generated by different methods. Participants are asked to select the avatar image that best matches the input expression without knowing the source of each image.
We use the Wilson confidence interval~\cite{yan2010stratified} to establish the confidence level for each response and analyze the result distributions. As shown in Fig.~\ref{pic:user_study}, high-confidence results (>=0.99) indicate that our method significantly surpasses Faceware (84\% vs. 5\%) and Animator (77\% vs. 21\%) in maintaining expression consistency. This substantiates our approach's effectiveness and practical utility in producing superior facial animation.

\begin{figure*}[t]
  \centering
  \includegraphics[width=0.98\linewidth]{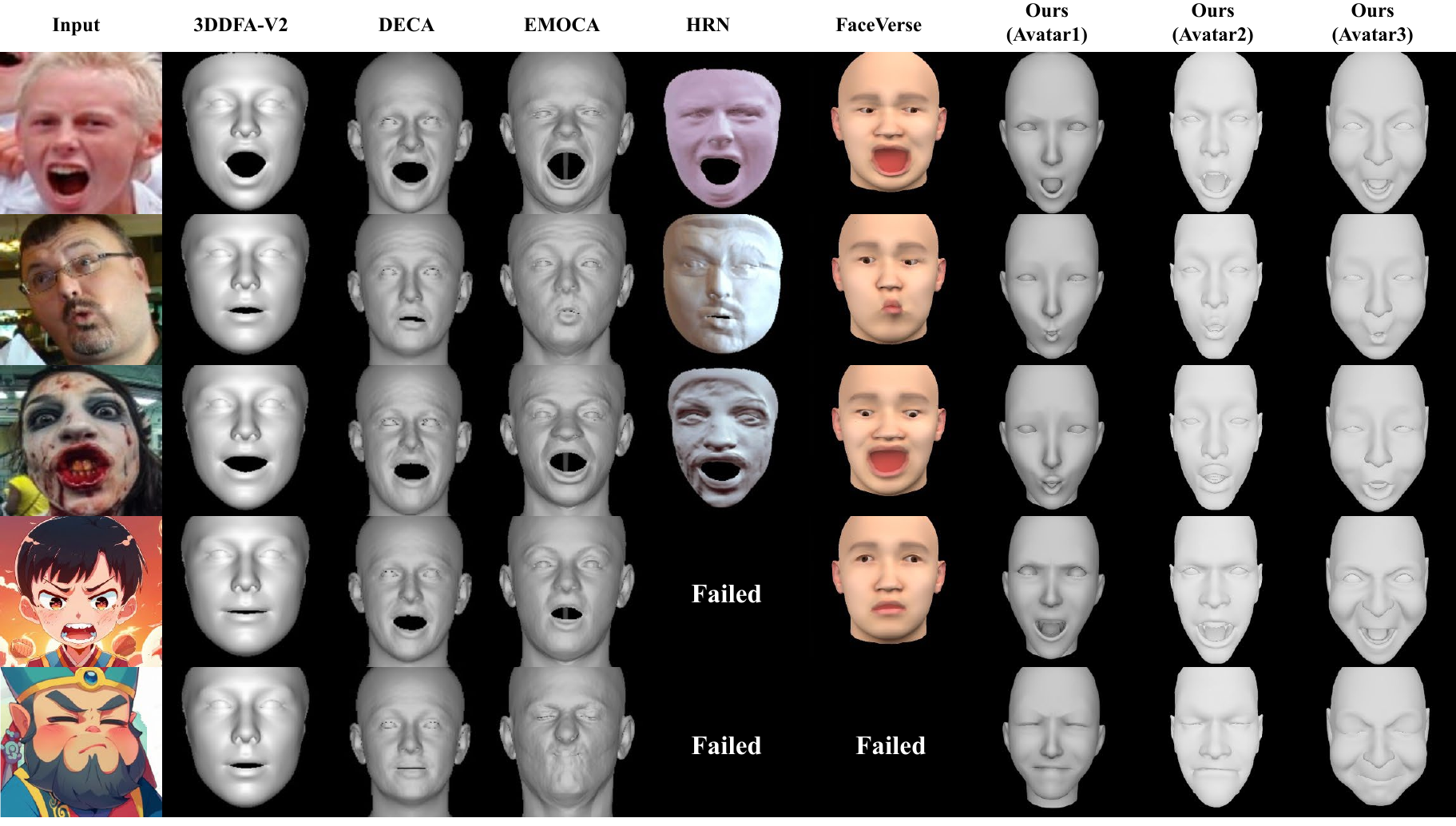}
  \caption{Comparisons with the state-of-the-art monocular face reconstruction methods in terms of maintaining consistent expressions. In some cases, HRN and FaceVerse fail to generate results due to the inability to detect facial landmarks. Portrait images are selected from FEC database \textcopyright{Google AI} (CC-0). \textit{Cartoon characters} \textcopyright{NetEase}. }
  \label{pic:exp_recons}
\end{figure*}

\subsection{Comparisons with Face Reconstruction Methods}
In this study, we compare our methods with three state-of-the-art monocular face reconstruction methods, i.e. 3DDFA-V2~\cite{guo2020towards}, DECA~\cite{feng2021learning}, EMOCA~\cite{danvevcek2022emoca}, HRN~\cite{lei2023hierarchical},FaceVerse~\cite{wang2022faceverse} in terms of maintaining consistent expressions when handling in-the-wild data.
These data exhibit a high level of diversity because they are not strictly constrained by factors such as head posture, background, occlusion, lighting conditions, and identity. This greatly simplifies and enhances the efficiency of the data collection process.

\subsubsection{Real footage.} We first test real footage with expressive emotions extracted from movies and TV shows. 
The results are shown in Fig.~\ref{pic:exp_recons}. Although HRN performs better in maintaining facial expression consistency compared to other facial reconstruction methods, factors such as occlusion and head posture can severely compromise its effectiveness. In contrast, our results more accurately convey the facial movements of the input images across different scenarios, including expressions such as smiling, screaming, and frowning. It is worth noting that our method still maintains better consistency in expressions, despite the differences in identity between the input and output, which could potentially complicate the process.



\subsubsection{Stylized cartoon characters.} We also conduct experiments on various stylized cartoon characters. These characters often have facial shapes and features that greatly differ from those of real humans. 
As shown in Fig.~\ref{pic:exp_recons}, facial reconstruction methods such as HRN and FaceVerse fail to generate face models in some cases. This is because these methods rely on the detection of facial landmarks. When facial landmarks cannot be detected, these methods do not work.
In contrast, our method still works and maintains the consistency of their facial expressions. We attribute this to the robust expressiveness and generalizability of our expression foundation model. These results suggest that our pipeline exhibits better performance and stronger applicability under real-world, non-ideal conditions.

\subsubsection{Multi-Avatars.} Additionally, our method is capable of training on multiple distinctively-rigged avatars simultaneously, without the need to train a separate network for each character as previous works required. In Fig.~\ref{pic:exp_recons}, we present the facial animation transfer results for three avatars with different appearances. Thanks to our identity-conditional semi-supervised training strategy and our foundational expression model, our method can adaptively generate consistent expression results for different avatars and their respective rigs. We also conducted user study to compare the effectiveness of multi-avatars and single-avatar decoders. The user study shows a comparable preference for the multi-avatar and single-avatar decoders with a ratio of 51.6\% to 48.4\%. This indicates that the strategy does not compromise the transfer quality. This method is particularly useful for multi-character inference in resource-limited scenarios, such as mobile games.

\subsection{Ablation Study}



\subsubsection{Ablation on expression representations.}
\begin{figure}[t]
  \centering
   \includegraphics[width=0.98\linewidth]{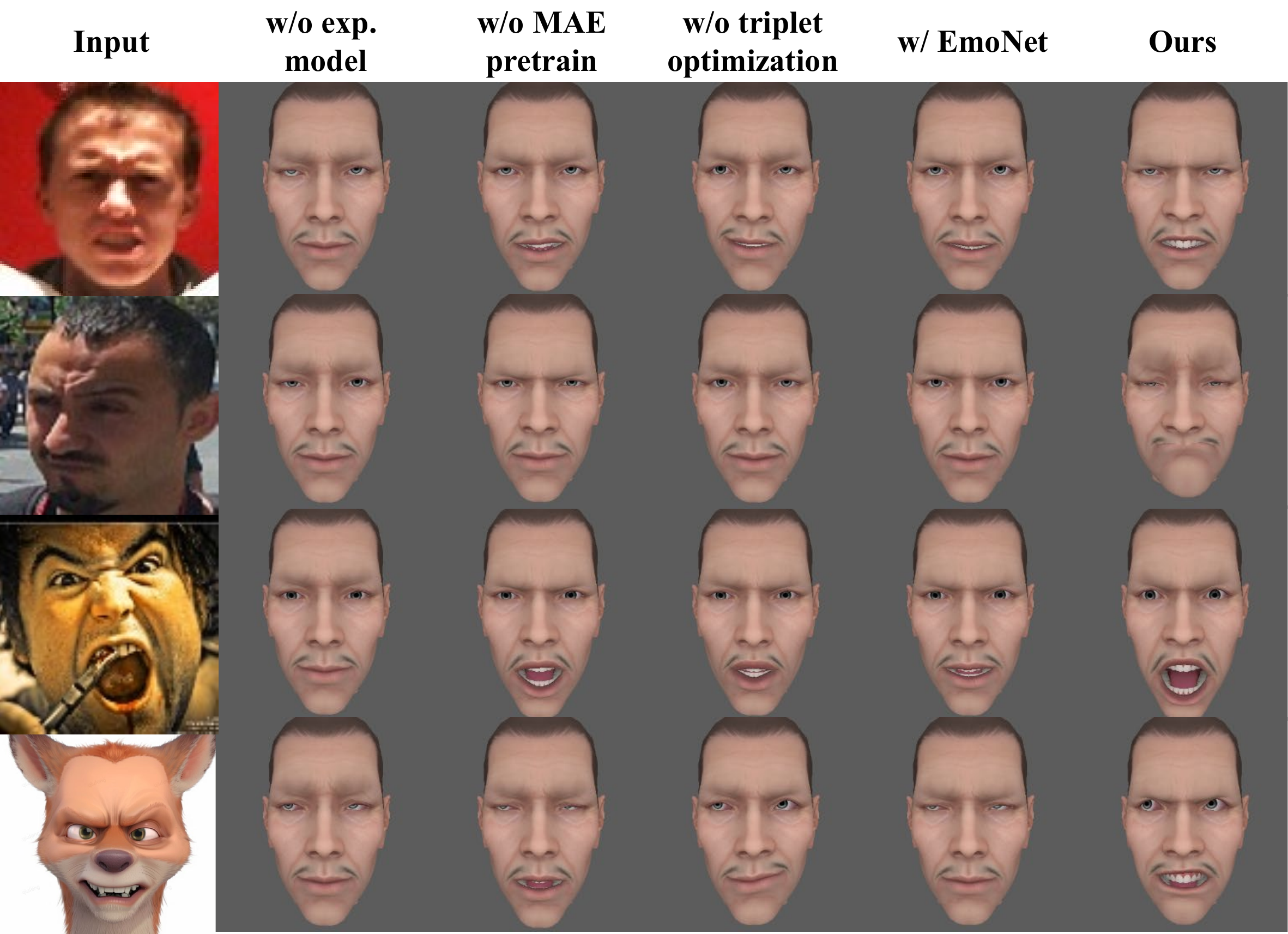}
  \caption{Comparison of different expression models. The results based on our expression foundation model contain more expression details. Portrait images are selected from FEC database \textcopyright{Google AI} (CC-0). \textit{Cartoon character} \textcopyright{NetEase}. }
  \label{pic:abla_expr}
\end{figure}

For facial animation transfer, it is crucial to explore an expression representation with sufficient capability and generalizability to constrain expression consistency. 
We first substitute our expression foundation model by randomly initializing the weights of the ViT encoder. As depicted in Fig.~\ref{pic:abla_expr}, it fails to capture the emotions from in-the-wild input images, which validates the necessity of the expression model. 

\begin{figure}[t]
  \centering
  \includegraphics[width=0.95\linewidth]{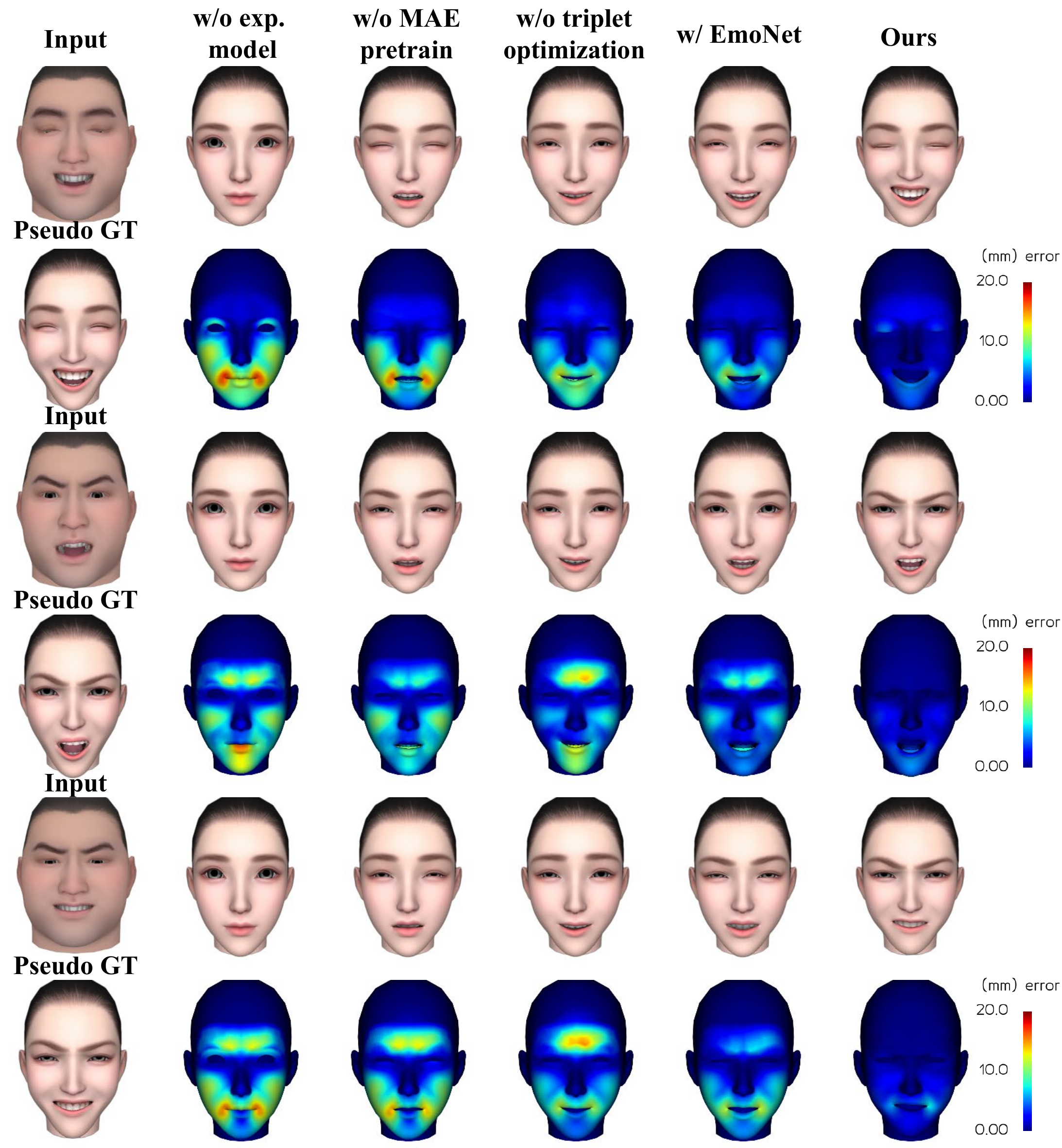}
  \caption{Visualizations of the mean squared error between the 3D meshes of generated results and pseudo-ground-truth. The red color represents higher error values, while the blue indicates lower errors.}
  \label{pic:abla_gt}
\end{figure}

We then explore the effectiveness of facial feature space reconstruction with MAE pretraining. Specifically, we train a ViT encoder with randomly initialized weights on the expression triplet datasets as a substitute for our expression foundation model. It is obvious that without MAE pretraining, the generated facial animations often fail to ensure consistency in expression, especially for stylized cartoon characters. 
This validates that MAE pretraining for facial auto-reconstruction can significantly enhance the generalizability of expression features, allowing for the extraction of semantically consistent expression features even from in-the-wild facial images.

Moreover, we validate the effectiveness of expression feature space optimization on our triplet dataset. Without triplet optimization, the accuracy of the ViT encoder in the expression comparison task drops significantly from 87.73\% to 34.06\%. 
In the animation transfer task, as depicted in the fourth column of Fig.~\ref{pic:abla_expr}, the transferred results fail to maintain expression consistency. This highlights the critical effort of the triplet optimization in accurate expression feature extraction.


We also investigate the difference between our expression foundation model and EmoNet~\cite{toisoul2021estimation}, which is the state-of-the-art emotion recognition model used in a facial reconstruction method EMOCA~\cite{danvevcek2022emoca} to capture 
emotions. EmoNet uses ResNet-50~\cite{he2016deep} as the backbone and is trained on an emotion recognition dataset AfffectNet~\cite{mollahosseini2017affectnet}. The features of the final layer are then employed as the emotional features. We replace the expression foundation model in our FreeAvatar with EmoNet, and the comparisons can be seen in Fig.~\ref{pic:abla_expr}. In most scenarios, although our framework with EmoNet can also achieve facial motion transfer, the fidelity of expression details is significantly lower than our FreeAvatar. 

For a more rigorous comparison, we evaluate the mean square errors between the 3D meshes of the generated results and the pseudo-ground-truth.
Specifically, we first deform the input character's 3D meshes to the target character's 3D meshes using deformation transfer~\cite{sumner2004deformation}. Since this is an optimization-based method, we treat the deformed target 3D meshes as the pseudo-ground-truth.  
Afterward, we animate the expression from the input character's image onto the target character via FreeAvatar.
Subsequently, we calculate the mean absolute error between the 3D meshes of the generated image and the pseudo-ground-truth. 
Visualizations in Fig.~\ref{pic:abla_gt} suggest that the 3d meshes of generated images have almost no deviation from the ground truths. 
This demonstrates our method has high precision and accuracy in capturing the expression details of the input images.

\begin{figure}[t]
  \centering
  \includegraphics[width=0.98\linewidth]{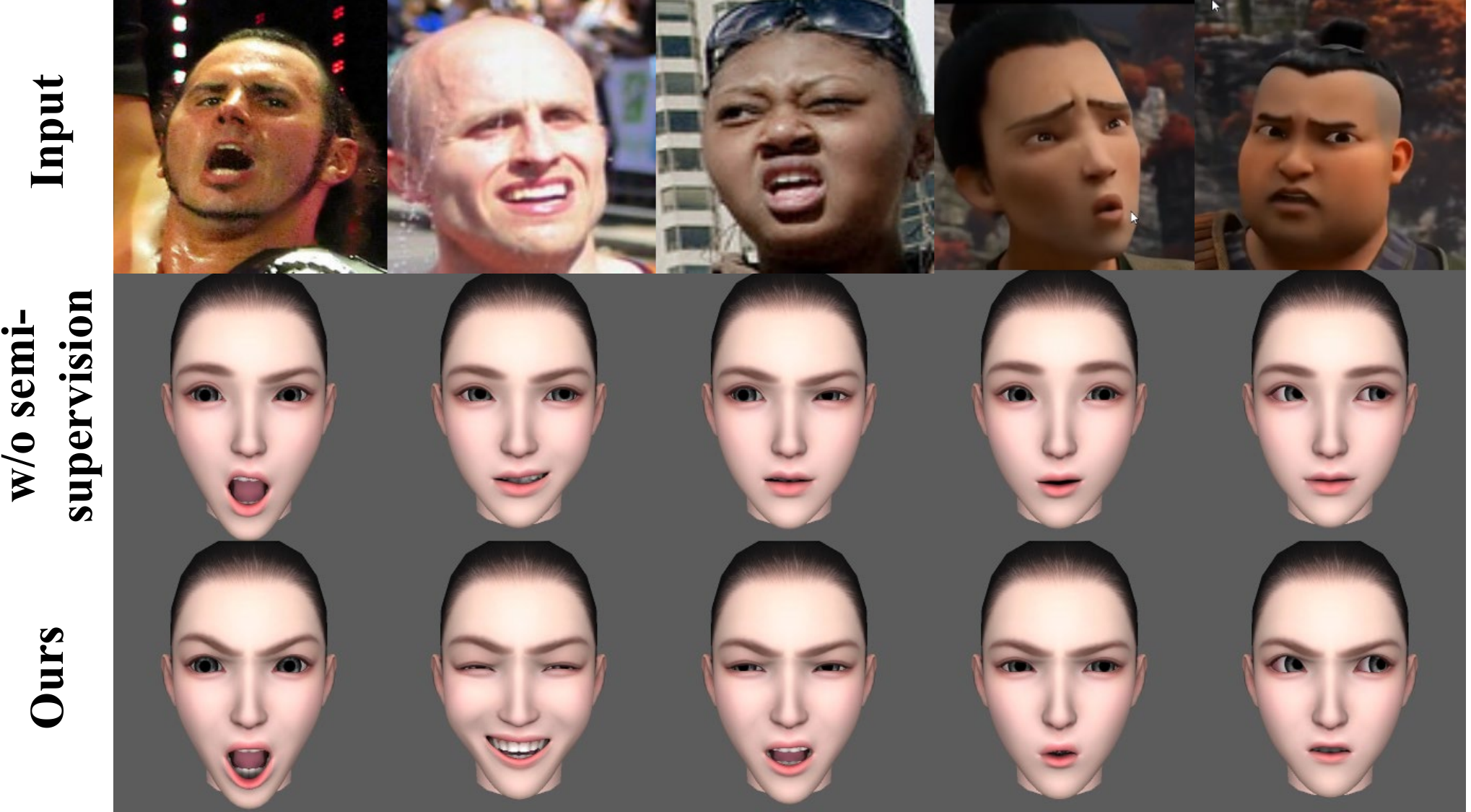}
  \caption{Ablation on semi-supervised learning. Benefiting from the superior generalization brought by semi-supervision, FreeAvatar adapts effectively to in-the-wild images. Portrait images are selected from FEC database \textcopyright{Google AI} (CC-0). \textit{Cartoon character} \textcopyright{NetEase}.}
  \label{pic:abla_weak}
\end{figure}
  
\begin{figure}[t]
  \centering
  \includegraphics[width=0.98\linewidth]{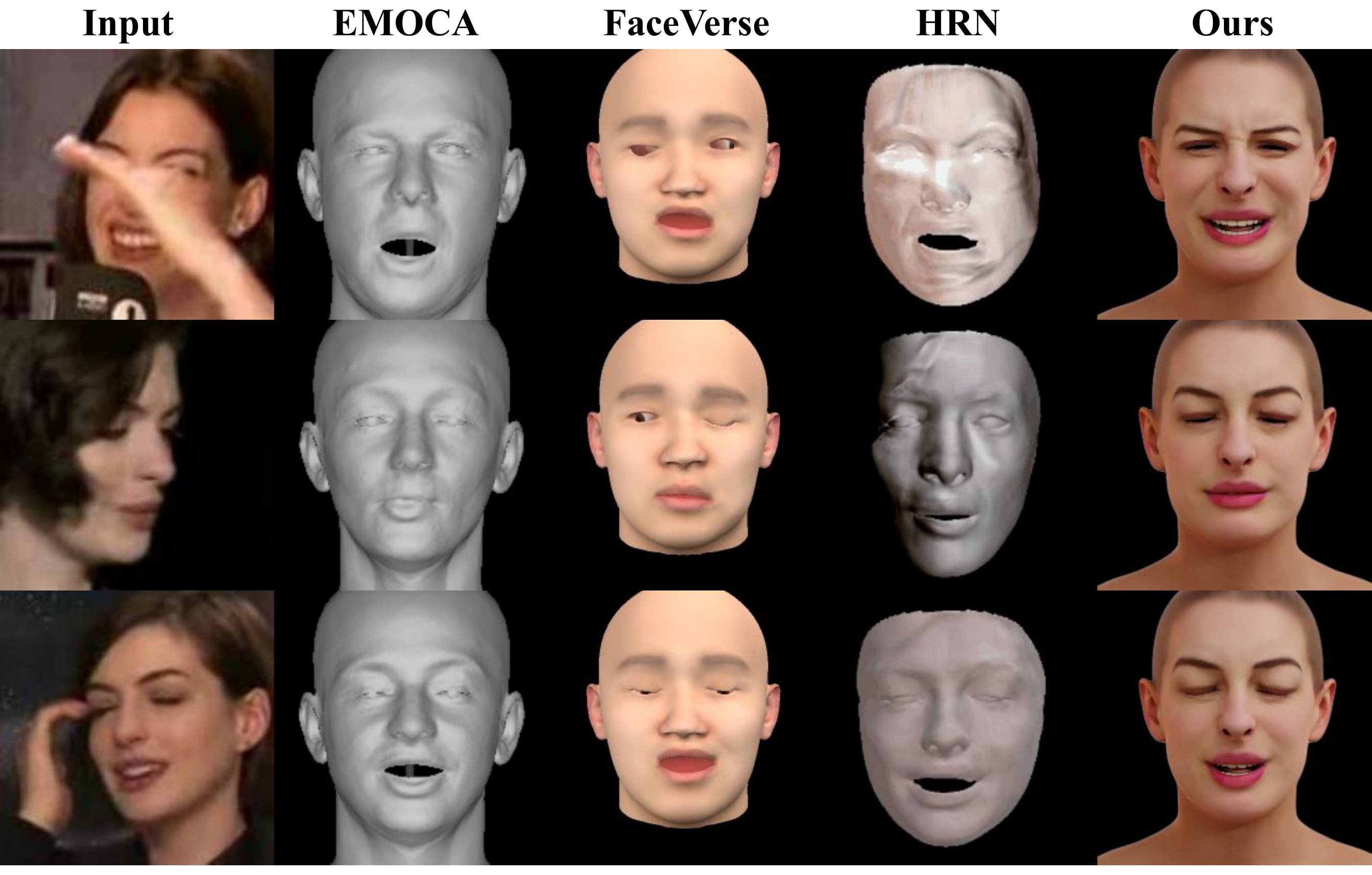}
  \caption{Comparison of actor-specific results. Our approach demonstrates superior stability under occlusions and variations in head poses. \textit{Anne Hathaway} images are selected from VoxCeleb~\cite{nagrani2017voxceleb} \textcopyright{Visual Geometry Group} (CC BY-SA 4.0).}
  \label{pic:self_recons}
\end{figure}
  

\subsubsection{Ablation on semi-supervised learning.}
As previously mentioned, our facial animation transfer module requires avatar data and a portion of in-the-wild data for training in a semi-supervised manner. In this study, we explore the effectiveness of semi-supervised learning for our FreeAvatar. Fig.~\ref{pic:abla_weak} shows a comparison between our results and those without semi-supervised learning. It can be observed that if we use only avatar data for fully supervised learning during training, the generalizability to in-the-wild test data is poor, making it challenging to generate facial animations with consistent expressions. This demonstrates that incorporating semi-supervised learning during training can significantly enhance the model's generalizability to in-the-wild test data.


\subsection{Exploration on Actor-specific Transfer}
Unlike facial reconstruction methods, FreeAvatar does not create head avatars but instead transfers facial animations to avatars with predefined rigs. 
Due to its versatility, we can seamlessly integrate FreeAvatar with facial reconstruction techniques to achieve actor-specific facial animation transfer.  
Specifically, we employ off-the-shelf models to create avatars resembling the input identities and then apply expression transfer with FreeAvatar.
As depicted in Fig.~\ref{pic:self_recons}, our method consistently maintains high fidelity and stability, even in challenging conditions such as occlusion and profile views. It demonstrates that our method is well-suited for a broader range of application scenarios.

\section{Discussion and Conclusion}
\subsection{Limitations}
Although FreeAvatar is capable of achieving high-quality facial animation transfer, it has certain limitations. 
First, it does not incorporate temporal information during training, relying instead on a post-processing strategy to mitigate face-jittering issues. 
Additionally, while our method demonstrates robustness against variations in lighting, occlusion, and profile views, it struggles with cases involving significant condition variations. 
In extreme situations, such as complete eye occlusion, the model can only approximate eye movements. 
Furthermore, the rig decoder and neural renderer require retraining whenever new avatars are introduced. 
We intend to address these challenges in our future work to improve the robustness and applicability of 3D Facial animation transfer models.

\subsection{Conclusion} This work describes FreeAvatar, a robust approach to 3D facial animation transfer that effectively utilizes a learned expression representation without relying on geometric constraints. FreeAvatar first develops an expression foundation model to construct a continuous and fine-grained expression feature space. Afterward, the Expression-driven Multi-avatar Animator encodes the extracted expression representation into facial rigs and imposes perceptual constraints to ensure expression consistency. To make this process differentiable, we employ a neural renderer to translate the rigs into corresponding facial images. Additionally, a dynamic identity injection module is designed to enable joint training for multiple avatars. Experimental results demonstrate FreeAvatar's superior performance and strong robustness. We hope that our method can enable many new possibilities for high-fidelity 3D facial animation transfer from in-the-wild faces.


%
%
%
%

\bibliographystyle{ACM-Reference-Format}
\bibliography{main}



\clearpage

\end{document}